\documentclass[aps,pre,twocolumn,nofootinbib,floatfix,showpacs]{revtex4}  

\usepackage{graphicx} 
\usepackage{amssymb}
\newcommand {\be}{\begin{equation}}
\newcommand {\ee}{\end{equation}}
\newcommand {\ba}{\begin{eqnarray}}
\newcommand {\ea}{\end{eqnarray}}

\begin{document}

\title{Size effects on thermal rectification in mass-graded anharmonic lattices}

\author{M.~Romero-Bastida and Armando~Gonz\'alez-Alarc\'on}
\affiliation{SEPI ESIME-Culhuac\'an, Instituto Polit\'ecnico Nacional, Avenida Santa Ana 1000, Colonia San Francisco Culhuac\'an, Delegaci\'on Coyoacan, Distrito Federal 04430, Mexico}
\email{mromerob@ipn.mx}

\date{\today}

\begin{abstract}
In this work we study the thermal rectification efficiency of a one-dimensional mass-graded anharmonic oscillator lattice at large system sizes. A modest increase in rectification is observed. When the magnitude of the mass gradient scales with the system size, the aforementioned effect increases substantially. This result can be unmistakeably attributed to an asymmetry in the local temperature profile obtained for the employed parameter values.
\end{abstract}


\maketitle

\section{Introduction\label{sec:Intro}}

A thermal rectifier is a device that has a high or infinite thermal resistance in one direction but a low resistance when the temperature gradient is reversed. From its first experimental observation to its latest implementation, there have been a whole array of mechanisms proposed to yield thermal rectification both at nanoscopic and macroscopic dimensions~\cite{Robertson11}. Recently there has been a large increase of theoretical studies concerning thermal rectification in nanoestructured materials that have shown promising results for the future construction of thermally rectifying devices such as thermal rectifiers, transistors, and memories~\cite{Li12}. Considering coupled anharmonic oscillator lattices, which are very convenient model systems to perform numerical studies, nearly all of the results so far reported consider lattices composed of two or three segments, each being a different nonlinear lattice~\cite{Terraneo02,Li04a,Hu05}; the rectifying mechanism corresponds to the shift of the effective phonon spectrum as temperature changes. All of the above thermal rectifiers will work well for small system sizes; however, as the latter increases, rectification will decrease dramatically~\cite{Hu06,Hu06a}.

Another inhomogeneous system with rectifying properties corresponds to an anharmonic lattice with nearest-neighbor interactions and a linear mass variation along its length~\cite{Yang07}, inspired by the first successful experimental implementation of a rectifying device employing nanotubes~\cite{Chang06a}. Rectification in this device has been shown to be robust against the variation of various structural parameters~\cite{Romero13} and it is also known that better rectification is obtained with an exponential mass variation~\cite{Shah12}. Afterward, the proposal of employing a graded property in the considered system has been further applied to a closed billiard model with a graded magnetic field along its length to obtain thermal rectification~\cite{Casati07} and also to a chain of elastically colliding, asymmetrically-shaped mass-graded particles, where rectification is theoretically predicted when the system has a local heat flux proportional to the temperature gradient and a local conductivity dependent on temperature~\cite{Wang12}. Recently, much theoretical insight has been obtained with the study of a graded harmonic lattice with a quartic on-site anharmonic potential, self consistent heat baths, and weak interparticle interactions~\cite{Pereira10b}. In a latter work it has been shown that, for this model, the conditions for the existence of thermal rectifications are the existence of a temperature gradient in the bulk, local conductivity dependent on temperature, and a graded structure~\cite{Pereira11}. Finally, when this model incorporates long-range harmonic interactions, an increase in rectification efficiency is obtained~\cite{Pereira13}. 

Despite the relevance of all the aforementioned results, the decrease of rectification efficiency with increasing system size remains a relevant problem in the field, and has been addressed with various degrees of success in all the previously cited works. For the billiard system with a graded magnetic field rectification is proportional to the inverse of the system size~\cite{Casati07}, and thus its use is restricted to small system sizes; the one with asymmetrically-shaped mass-graded particles has a rectification that is largely independent of the system size $N$~\cite{Wang12}. For the harmonic oscillator chain with an on-site anharmonic potential, which takes into account the unavoidable dissipation in the bulk of the system, it was first noticed that, if the graded mass grows as $N^2$, rectification will not decay with $N$~\cite{Pereira10b}. Later on it was determined that an exponential mass distribution is needed to prevent the loss of rectification with increasing $N$~\cite{Pereira11}. Furthermore, an increase of rectification is attained if harmonic long-range interactions with slow polynomial decay are considered~\cite{Pereira13}; however, these latter type of interactions are difficult to obtain.

For the original mass-graded anharmonic lattice, which has received much less attention in the literature, a common feature of the first studies is that considered system sizes are rather small~\cite{Yang07,Shah12}. For a linear mass variation it has been reported that, in the case of small mass gradients, a system size of 400 oscillators has better rectification figures, although only marginally above those for the original case of 200 oscillators~\cite{Romero13}. This latter result hints that there might be a relation between the system size and the magnitude of the mass loading, but that so far has not been explored as a possible mechanism to improve the rectification power of this type of system.

In this work we propose to explore more systematically the aforementioned relation between system size and mass loading both to gain a deeper understanding of the rectification of mass-graded anharmonic lattices and to improve its efficiency. We propose that by properly scaling the magnitude of the mass-loading with the system size, a significant increase of rectification is obtained for large system sizes so far not studied. A mechanism responsible of rectification is observed that seems to be related to a peculiar nonequilibrium steady state that arises as a result of the employed system parameters. 

This paper is organized as follows: In Sec.~\ref{sec:Model} the model system and methodology are presented. Our results, obtained with the above mentioned scaling that increases rectification at large system sizes, are reported in Sec.~\ref{sec:Res}. The discussion of the results, as well as our conclusions, are presented in Sec.~\ref{sec:Disc}.

\section{system description\label{sec:Model}}

The reference Hamiltonian for the one dimensional model we are considering can be written as $H=\sum_{i=0}^{\mathcal N}H_i=\sum_{i=0}^{\mathcal N}\left[p_i^2/2m_i + V(q_{i+1}-q_i)\right]$, where ${\mathcal N}$ is the system size and $V(x)$ is the nearest-neighbor interaction potential. $\{m_i,q_i,p_i\}_{i=1}^{\mathcal N}$ are the dimensionless mass, displacement, and momentum of the $i$th oscillator; fixed boundary conditions are assumed ($q_{_0}=q_{_{{\mathcal N}+1}}=0$). The hereafter considered explicit form for the nearest-neighbor potential corresponds to the Fermi-Pasta-Ulam (FPU) $\beta$ model~\cite{FPU}, characterized by $V(x) = x^2/2 + x^4/4$. We consider a mass-graded lattice, wherein the mass of the {\sl i}th oscillator is given by $m_i=M_{\mathrm{max}}-(i-1)(M_{\mathrm{max}}-M_{\mathrm{min}})/({\mathcal N}-1)$. $M_{\mathrm{max}}$ and $M_{\mathrm{min}}$ are the mass of the oscillators at left and right ends, respectively, which define the mass gradient as $\Delta M\equiv M_{\mathrm{max}}-M_{\mathrm{min}}$. In the following we will always take $M_{\mathrm{min}}=1$. The equations of motion (EOMs) for each lattice oscillator can be written as $\dot q_i =p_i/m_i$ and
\ba
\dot p_i&=&F(q_i-q_{i-1})-F(q_{i+1}-q_i) \cr
   & + & \sum_{j=1}^{n_{_L}}(\xi_{_L}^{(i)} - \lambda_{_L} p_i)\,\delta_{ij} \cr
   & + & \sum_{k={\mathcal N}-n_{_R}+1}^{\mathcal N} (\xi_{_R}^{(i)} - \lambda_{_R} p_i)\,\delta_{ik},
\ea
where $F(x)=-\partial_x V(x)$ and $\xi_{_{L/R}}^{(i)}$ is a Gaussian white noise with zero mean and correlation $\langle\xi_{_{L,R}}^{(i)}(t)\xi_{_{L,R}}^{(j)}(t^{\prime})\rangle=2\lambda_{_{L,R}}k_{_B}T_{_{L,R}}\delta_{ij}\delta(t-t^{\prime})$, being $\lambda_{_{L,R}}$ (taken as $=0.1$ in all computations hereafter reported) the coupling strength between the system and the left and right thermal reservoirs operating at temperatures $T_{_L}$ and $T_{_R}$, respectively; furthermore, $T_i=T_{_L}$ for $1\le i\le n_{_L}$ and $T_i=T_{_R}$ for ${\mathcal N}-n_{_R}+1\le i\le{\mathcal N}$. The aforementioned EOMs were integrated with a stochastic velocity-Verlet integrator with a timestep of $\Delta t=0.01$ for $10^{11}$ time units after a transient of $10^7$ time units.

Once the non-equilibrium stationary state is attained, the mean heat flux is computed as $J(t)= N^{-1}\sum_{i=n_{_L}+1}^{{\mathcal N}-n_{_R}}\dot q_{i} F(q_{i+1}-q_i)$, where $N$ is the number of unthermostated oscillators. Then the total heat flux is obtained as $J=\langle J(t)\rangle$, where $\langle\cdots\rangle$ indicates time average. We use $J_{+}$ to denote the heat flux obtained when the high temperature reservoir is attached to the heavy mass end and $J_{-}$ when the hot reservoir is attached to the light mass end. The \emph{rectifying efficiency} can thus be computed from the expression first proposed in Ref.~\cite{Eckmann06} that reads as
\be
r(J_+,J_-)\equiv r={\max\{|J_{+}|,|J_{-}|\}\over\min\{|J_{+}|,|J_{-}|\}}.
\ee
In the following the behavior of this quantity will be studied for constant values of the temperature gradient $\Delta T=T_{_L}-T_{_R}=0.16$ and the average temperature $T_{_0}\equiv(T_{_L}+T_{_R})/2=0.1$ of the system.

\section{Results\label{sec:Res}}

\begin{figure}
\includegraphics[width=0.99\linewidth,angle=0.0]{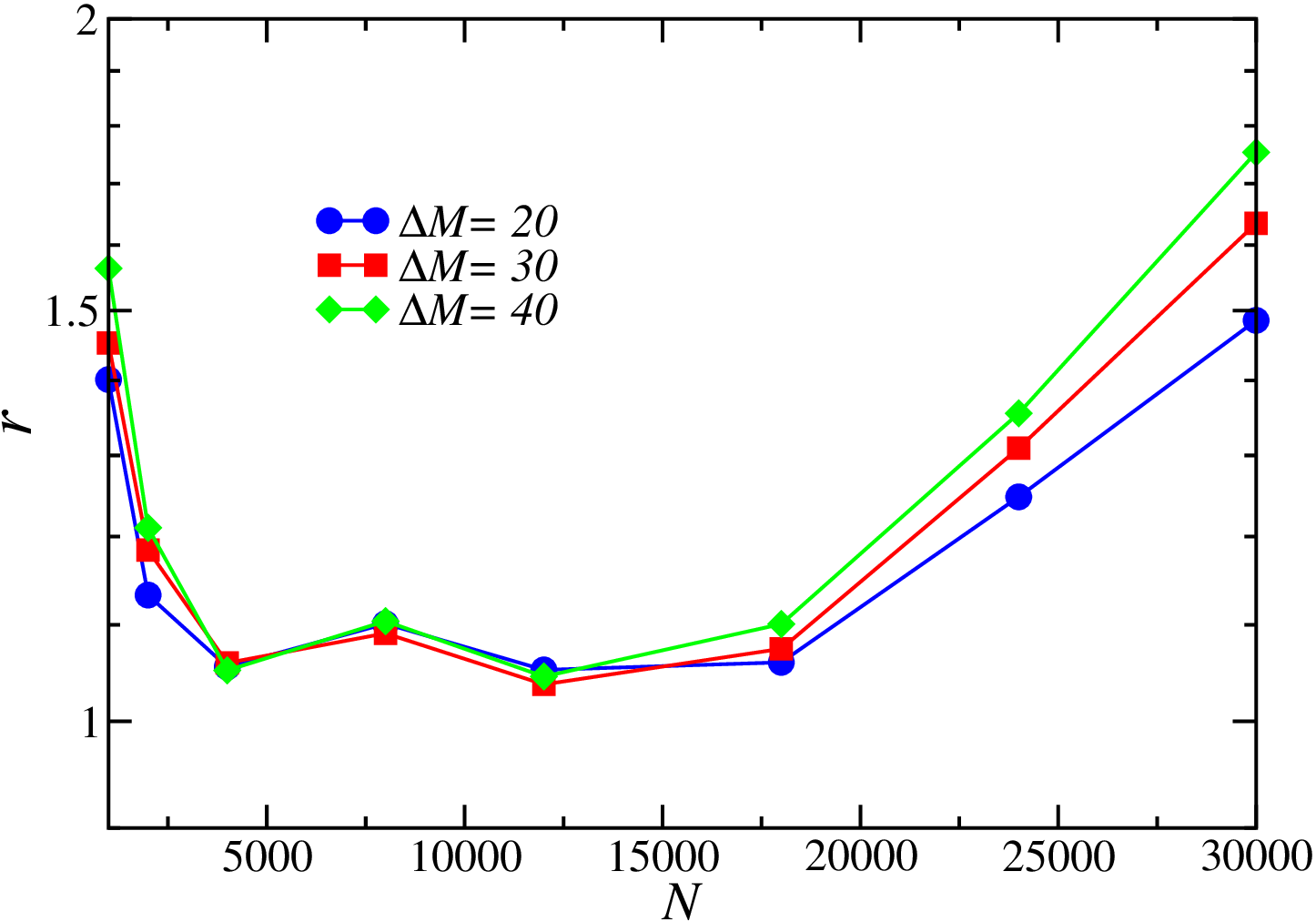}
\caption{(Color online) Thermal rectification $r=r(J_{+},J_{-})$ as a function of the system size $N$ for a mass gradient of $\Delta M=20$ (circles), $30$ (squares), and $40$ (diamonds), with $T_{_0}=0.1$, $\Delta T=0.16$, and $n_{_L}=n_{_R}=3$.}
\label{fig:uno}
\end{figure}

In Fig.~\ref{fig:uno} we plot the thermal rectification as a function of the system size for various mass gradients and a number of thermostated oscillators of $n_{_L}=n_{_R}=3$. We readily observe that, for $N<4000$, rectification decreases, in accordance to the results of previous work for other systems~\cite{Hu06,Hu06a}. For system sizes in the interval $4000<N<18\,000$ rectification remains largely constant. However, we notice that, as the system size further increases, a small but steady increase of rectification efficiency is clear for all considered mass gradients. Furthermore, although the variations are rather small, it can be appreciated that rectification presents a weak dependence on the mass gradient value for the largest system size considered ($N=30\,000$). Also, it is worth mentioning that, for this same case, the obtained rectification values are in the range $1.48<r<1.75$, which are very close to the figure of $1.43$ obtained experimentally in a device made of two cobalt oxides with different thermal conductivities in the low-temperature regime~\cite{Kobayashi09}.

\begin{figure}
\includegraphics[width=0.95\linewidth,angle=0.0]{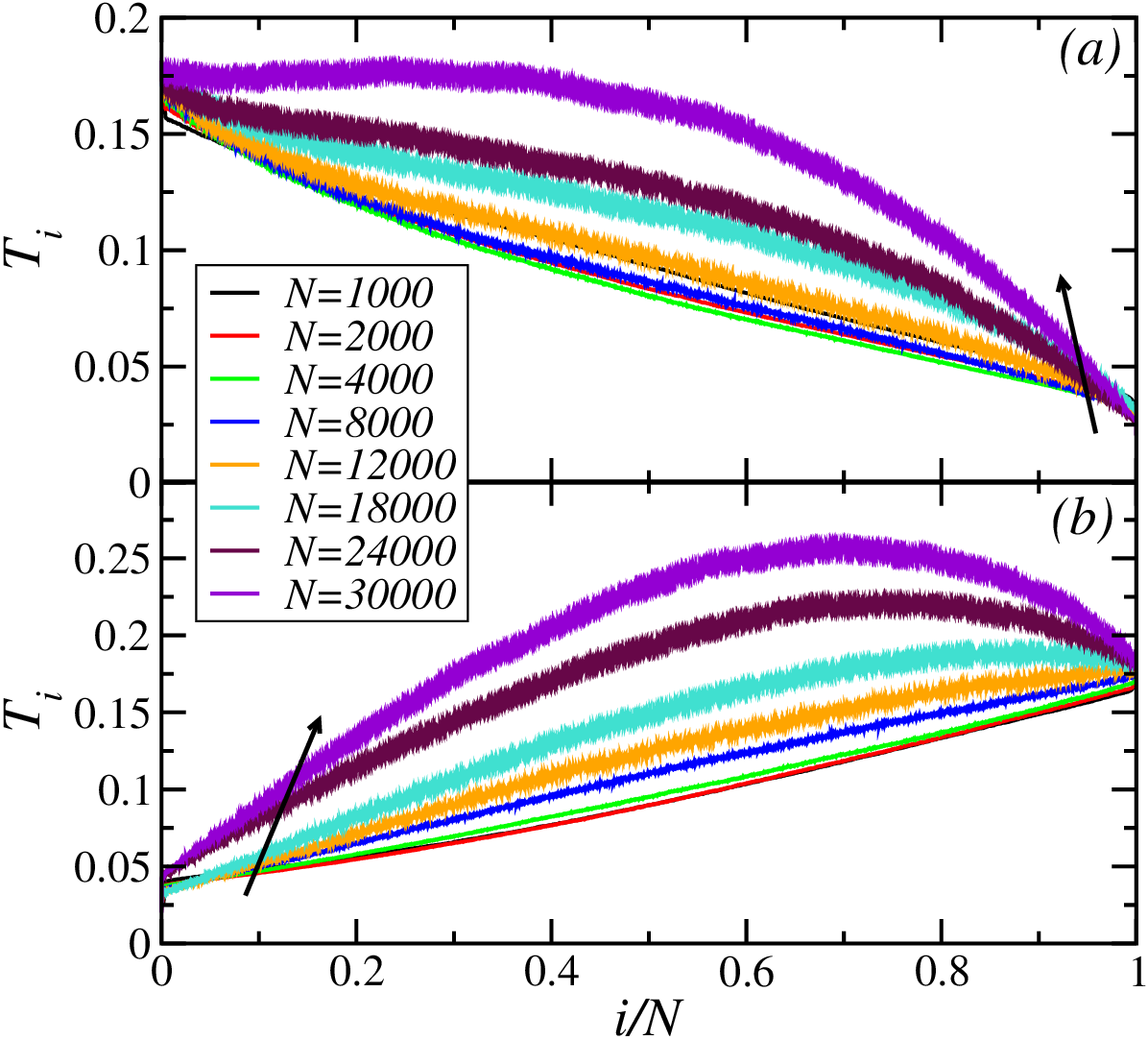}
\caption{(Color online) Temperature profiles for the $\Delta M=20$ case depicted in Fig.~\ref{fig:uno} using the same $T_{0}$, $\Delta T$, and $n_{_{L,R}}$ values as in previous figure. In panel (a) the case corresponding to $T_{_L}>T_{_R}$ is shown, whereas in panel (b) is the case when the two reservoirs at the ends are swapped, i.e. $T_{_L}<T_{_R}$. Arrows in both figures indicate the direction in which system size increases.}
\label{fig:dos}
\end{figure}

In order to explain the increase in the $r$ value for large system sizes, we plot in Fig.~\ref{fig:dos} the spatial profile of the local temperature $T_i=\langle p_i^2/m_i\rangle$ for the $\Delta M=20$ case considered in Fig.~\ref{fig:uno}. It is evidently a drastic departure from the well defined steady state characteristic of small sizes reported in Ref~\cite{Yang07}, albeit for a harmonic mass-graded lattice, since now the details of the temperature profile (TP) depend on both system size and mass gradient. For the largest system size depicted, which shows the greater departure from the aforementioned small-size steady state, we computed the first three even moments of the momentum and confirmed, by their agreement, that we have a Gaussian momentum distribution and thus a well defined local temperature. Furthermore, we have also corroborated that $\langle dH_i(t)/dt\rangle\sim10^{-7}$ for all oscillators. Thus a steady state, whose existence was proved in Ref.~\cite{Eckmann99a} for anharmonic lattices, has been indeed attained.

\begin{figure}
\includegraphics[width=0.99\linewidth,angle=0.0]{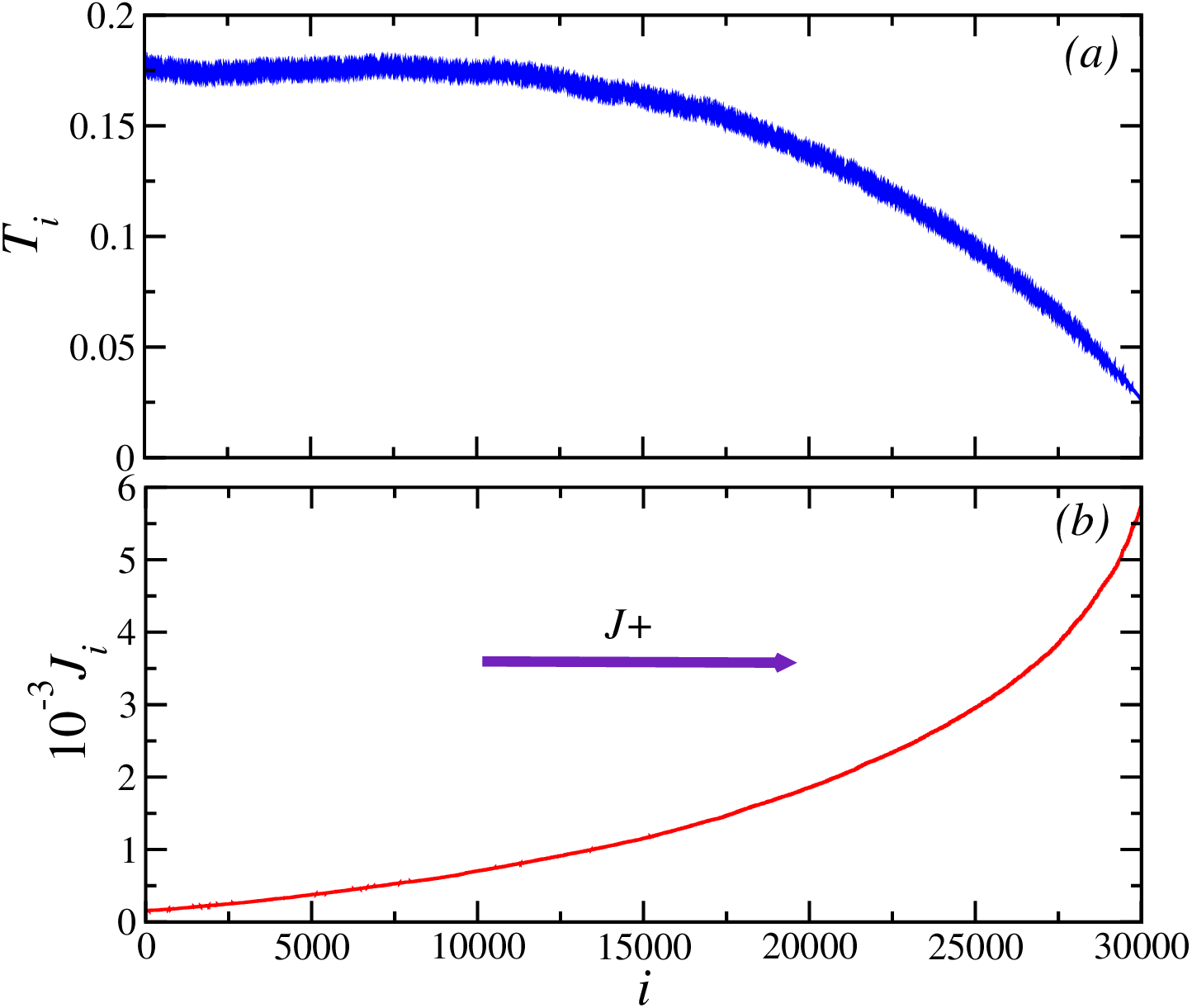}
\caption{(Color online) (a) Temperature profile for the $N=30\,000$ case depicted in Fig.~\ref{fig:dos}(a); same $T_{0}$, $\Delta T$, and $\Delta M$ values as in previous figure. (b) Corresponding local heat flux $J_{i}$ to the temperature profile in panel (a). Arrow in panel (b) indicates the direction of the total heat flux $J_{+}$.}
\label{fig:tresab}
\end{figure}

\begin{figure}
\includegraphics[width=0.99\linewidth,angle=0.0]{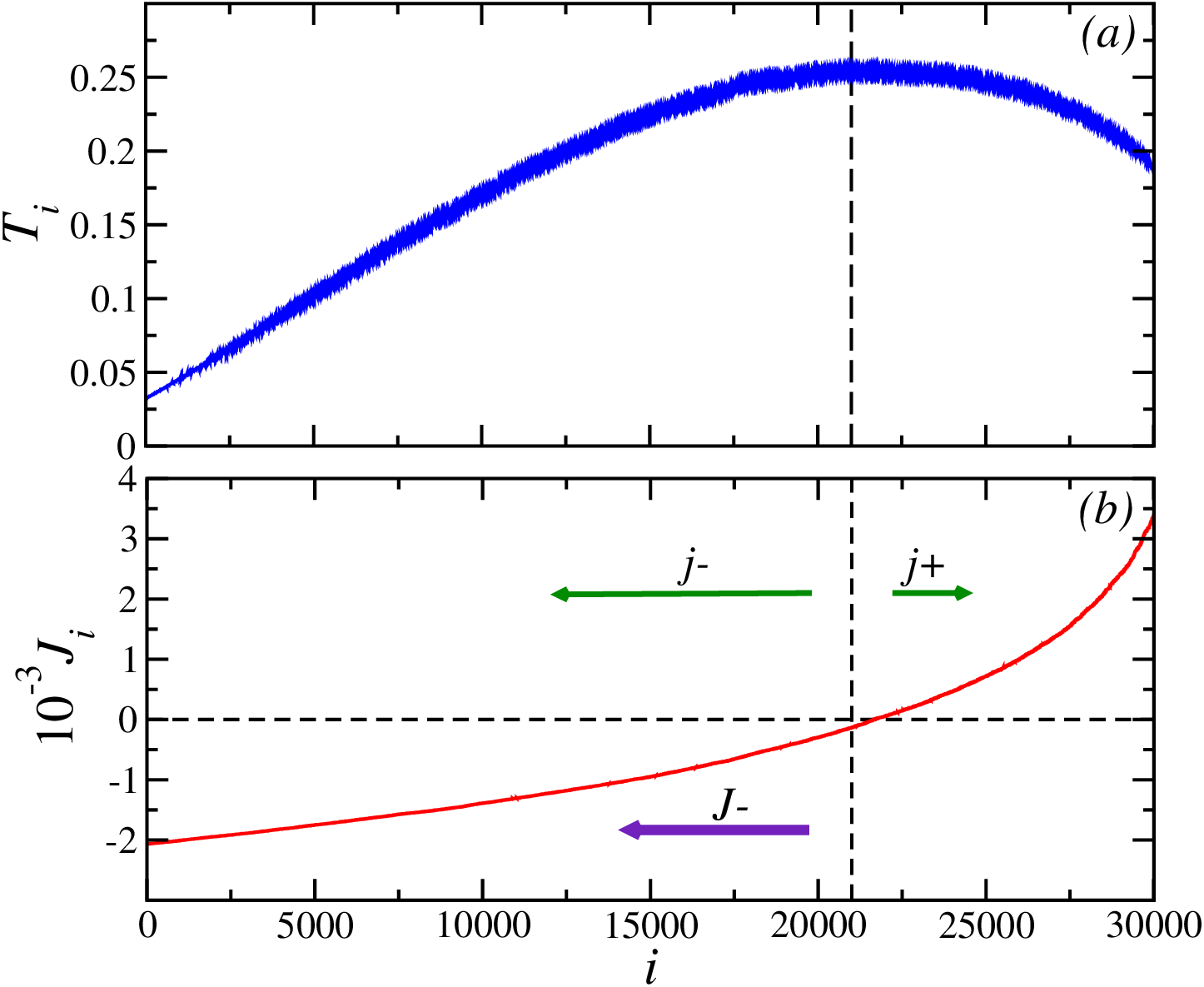}
\caption{(Color online) (a) Temperature profile for the $N=30\,000$ case depicted in Fig.~\ref{fig:dos}(b). (b) Corresponding local heat flux $J_{i}$ to the temperature profile in (a). Thick arrow in (b) indicate the direction of the total heat flux $J_{-}$ and thin arrows those of the ``local" heat fluxes $j_{+}$ and $j_{-}$ (see text for more details). In both figures the vertical dashed line indicates the position along the lattice length wherein the TP attains its maximum value.}
\label{fig:cuatroab}
\end{figure}

In Figs.~\ref{fig:tresab} and~\ref{fig:cuatroab} we isolate the TP for the $N=30\,000$ cases presented in Fig.~\ref{fig:dos} to study in greater detail the heat transport for this system size. The TP corresponding to $J_{+}$ is plotted in Fig.~\ref{fig:tresab}(a) and it can be readily seen that the slope does not change sign along the system length; therefore, the local heat flux $J_{i}$, presented in Fig.~\ref{fig:tresab}(b), is positive along the whole system length and thus $J_{+}$ properly describes the energy transport for the corresponding configuration of the thermal reservoirs. However, it is also worth noticing that, although monotonic, the TP has no constant slope; therefore, it could naturally be expected that the local heat flux would present some structure, as can indeed be observed again in Fig.~\ref{fig:tresab}(b), where $J_{i}$ slowly increases from the heavy-loaded end to the lighter one; this increment closely correlates with that of the slope in the TP. It is worth mentioning that an increase, albeit linear, of the local heat flux has been reported for a FPU lattice with a Frenkel-Kontorova (FK) on-site potential and a small temperature gradient, just as in the herein considered case~\cite{Zhao09}. Now, when the thermal reservoirs are swapped, the TP, presented in Fig.~\ref{fig:cuatroab}(a), displays an unusual and interesting feature: Its slope does change sign within the system's bulk, with a central region at a higher temperature than either heat reservoir; although uncommon, such TPs have indeed been previously reported in the literature for similar systems~\cite{Nianbei09,Iacobucci11}. Since the sign of the average slope within a spatial region in the temperature profile determines the direction of the associated heat flux, we can infer the existence of two opposing heat fluxes associated with the regions of average positive and negative slopes in the TP clearly visible in Fig.~\ref{fig:cuatroab}(a). This can indeed be confirmed in Fig.~\ref{fig:cuatroab}(b), where it is clearly seen that the local heat flux is positive and negative in the spatial regions corresponding to negative and positive slopes of the TP in Fig.~\ref{fig:cuatroab}(a). Therefore, it is possible to define average heat fluxes in these spatial regions that can be estimated as $j_+=10^{-3}$ and $j_-=-1.6\times10^{-3}$, corresponding to the averages over the spatial regions wherein the TP has positive and negative slopes, respectively. It is also worth remarking that the possible existence of such seemingly counterintuitive heat fluxes for anharmonic oscillator lattices has already been conjectured for some time~\cite{Eckmann04}. We observe that $j_{-}$ has the same order of magnitude that $J_{-}=-1.1\times10^{-3}$, and therefore can approximately describe the energy transport in the bulk. Now, since $J_+=1.6\times10^{-3}$, it is clear that, without the $j_+$ contribution, we would have $r(J_+,j_-)=1$, i.e. no rectification altogether. Therefore, the overall effect of $j_+$ is to render a modest increase in the rectification efficiency, i.e. $r(J_+,J_-)=1.5$ and already noticed in Fig.~\ref{fig:uno}.

As already mentioned, it can be observed in Fig.~\ref{fig:uno} that there is a small increment in $r$, from $1.48$ to $1.75$, for the largest lattice size reported, as the mass gradient does so. This result is an unexpected one because a natural guess would be that the effect of the mass gradient diminishes at increasing system sizes. On the contrary, its effect, i.e. greater rectification, is stronger for the largest system size reported. Therefore, a reasonable assumption is that, if the effect of the mass gradient can be somehow increased in the same proportion as the system size increases, rectification will become significant at large system sizes. Thus we propose to scale the effect of $\Delta M$ by considering the ``mass-gradient density" $\rho\equiv\Delta M/N$ as the proper parameter to control the influence of both the system size and mass gradient in a consistent way. This choice will keep the effects of the mass gradient responsible for the increase in rectification, as displayed in the behavior of the temperature profile, as $N$ increases.

\begin{figure}
\includegraphics[width=0.99\linewidth,angle=0.0]{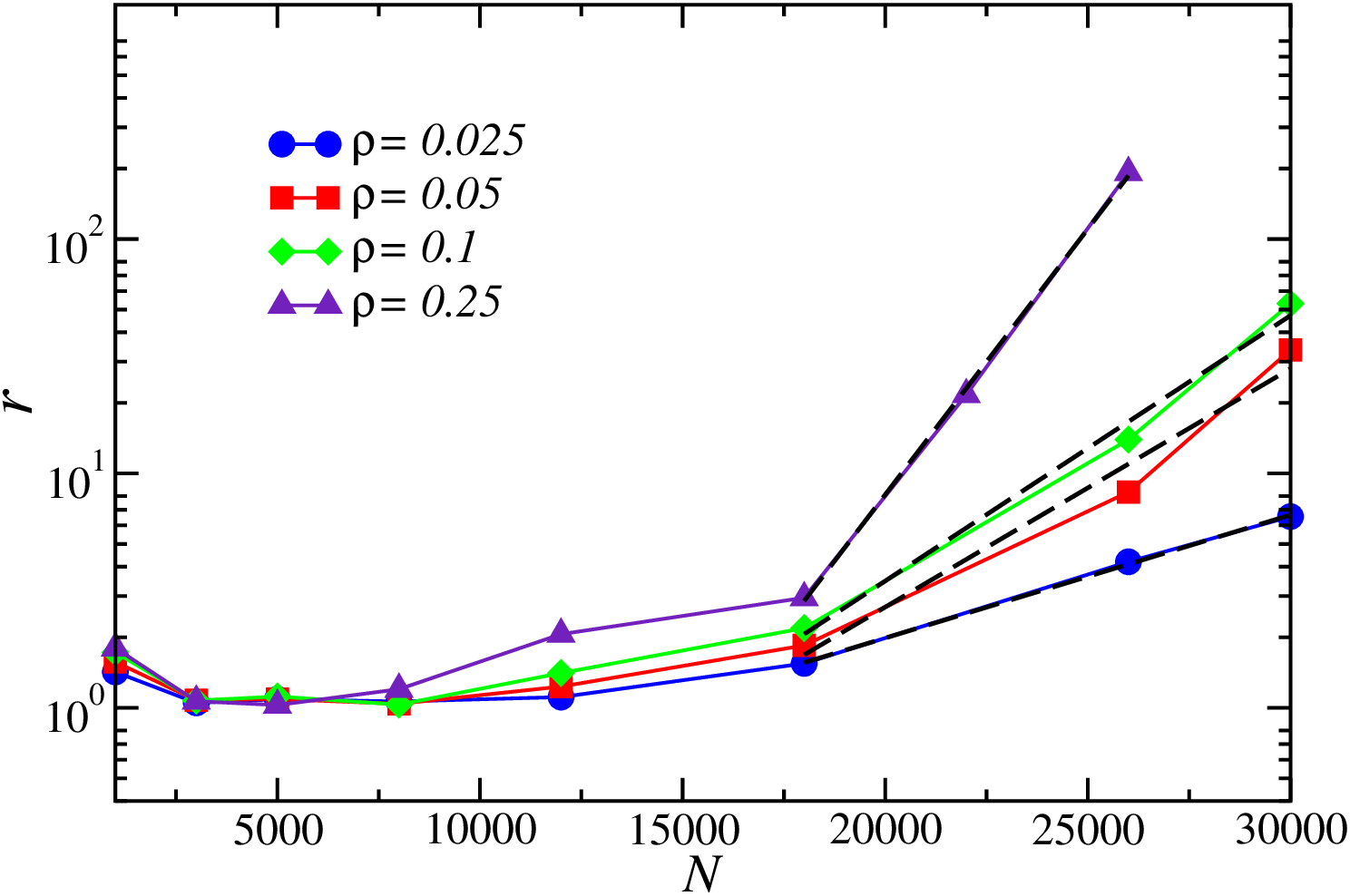}
\caption{(Color online) Thermal rectification $r$ vs system size $N$ for $\rho=\Delta M /N$ values of $0.025$ (circles), $0.05$ (squares), $0.1$ (diamonds), and $0.25$ (triangle up). Same $T_{0}$ and $\Delta T$ values as in Fig.~\ref{fig:uno}. Dashed lines indicate the corresponding exponential fit (see text for details).}
\label{fig:tres}
\end{figure}

In Fig.~\ref{fig:tres} we plot again the rectification vs system size, but now for different $\rho$ values. Striking differences are evident when compared to the results of Fig.~\ref{fig:uno}, since $r$ now has a steady increase up to figures of the order $10^2$ for increasing $\rho$ values. More precisely, for system sizes $N>18\,000$ rectification has an exponential dependence $r\sim\exp(\alpha N)$, where $\alpha\sim 2.5\times 10^{-4}$ for $0.05<\rho<0.15$. However, for larger $\rho$ values $\alpha$ increases, i.e. $\alpha=5.2\times10^{-4}$ for $\rho=0.25$. Therefore, for this last value a rectification of the order of $10^8$ can be easily obtained for $N=40\,000$. Thus, with a seemingly small change of variable we have been able to make the efficiency of the system approach the ideal limit for large $N$ values.

\begin{figure}
\includegraphics[width=0.99\linewidth,angle=0.0]{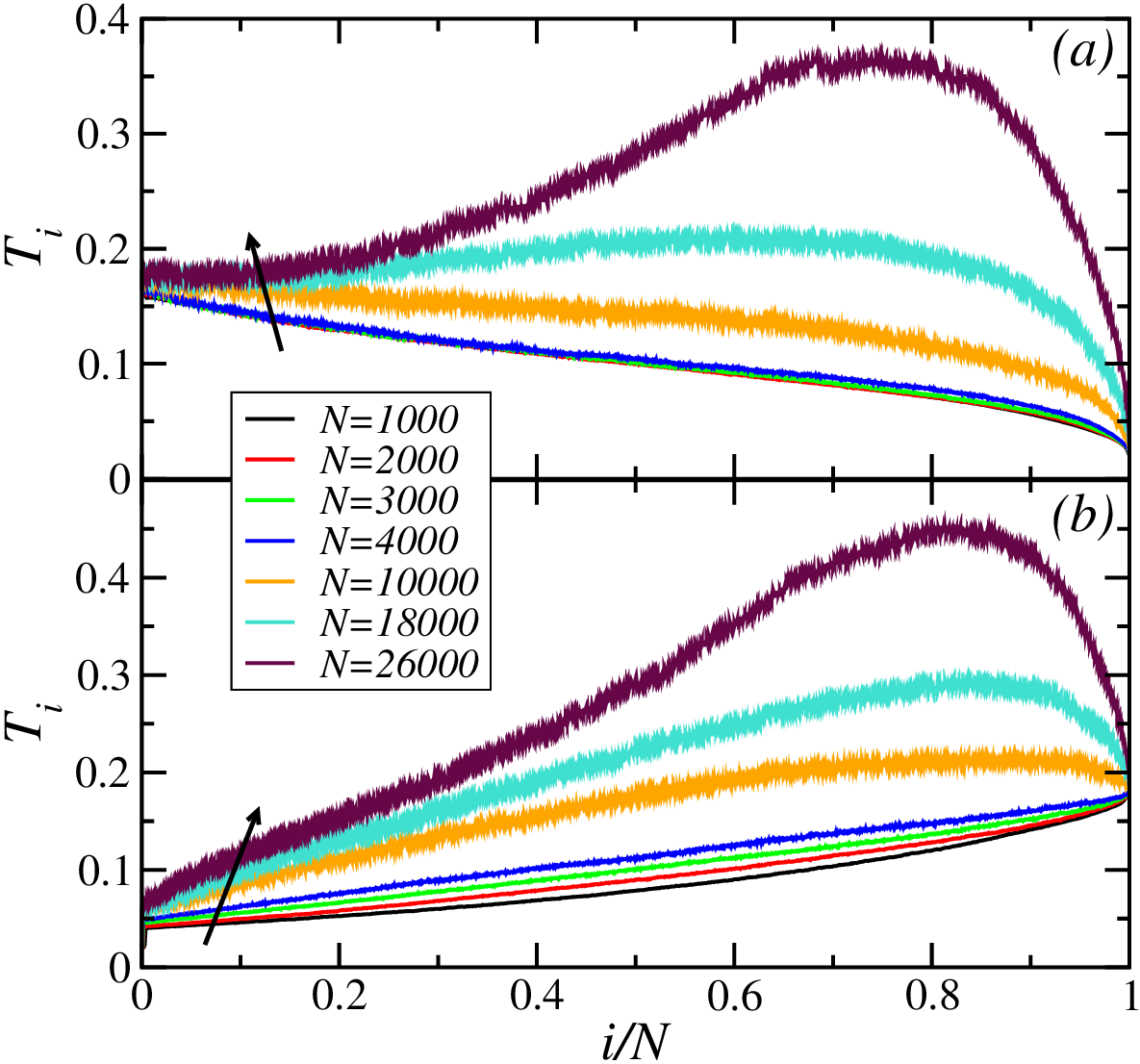}
\caption{(Color online) Temperature profiles for cases (a) $J_{+}$ and (b) $J_{-}$, both with $\rho=0.25$; the same $n_{_{R,L}}$, $T_{0}$, and $\Delta T$ values are used as in Fig.~\ref{fig:uno}.}
\label{fig:cuatro}
\end{figure}

Next we consider the TP for the mass-gradient density value that renders the higher $r$ value in Fig.~\ref{fig:cuatro}. The same phenomenology observed in Fig.~\ref{fig:dos} is herein obtained, albeit with the features responsible for the $r$ increase more pronounced. It is to be noted that for the $J_+$ case depicted in Fig.~\ref{fig:cuatro}(a) the TP is asymmetrical, just as the $J_-$ case depicted in Fig.~\ref{fig:dos}(b). Nevertheless, from the figure it can be observed that the positive and negative slopes, and thus the corresponding average heat fluxes, have different magnitudes; therefore, it is reasonable to expect a different total heat flux value in each case and, if their difference in magnitude is high enough, a significant rectification value. Explicitly, the total heat fluxes in Figs.~\ref{fig:cuatro}(a) and \ref{fig:cuatro}(b) are $J_+=10^{-4}$ and $J_-=-5.2\times10^{-7}$, respectively. Although the magnitudes are lower than those obtained for the case depicted in Fig.~\ref{fig:dos}, we now have two quantities that have a difference of three orders of magnitude, thus increasing the rectification with respect to that obtained with a constant $\Delta M$. Now, for the $J_+$ case the corresponding average heat fluxes are $j_+=5\times10^{-4}$ and $j_-=-5\times10^{-6}$. Thus $j_+$ dominates and there is energy transport in the direction of the decreasing mass gradient. On the other hand, for the temperature profile corresponding to the swapped heat bath configuration depicted in Fig.~\ref{fig:cuatro}(b) we have $j_+=4.7\times10^{-4}$ and $j_-=-1.2\times10^{-4}$. In this latter case both fluxes are of the same order of magnitude and thus neutralize each other. The total heat flux is thus impaired in the direction of increasing mass, as quantified by the low $J_-$ value reported above. This is the same mechanism already observed in Fig.~\ref{fig:dos}, just with a change in the scale of the involved fluxes.

\begin{figure}
\includegraphics[width=0.99\linewidth,angle=0.0]{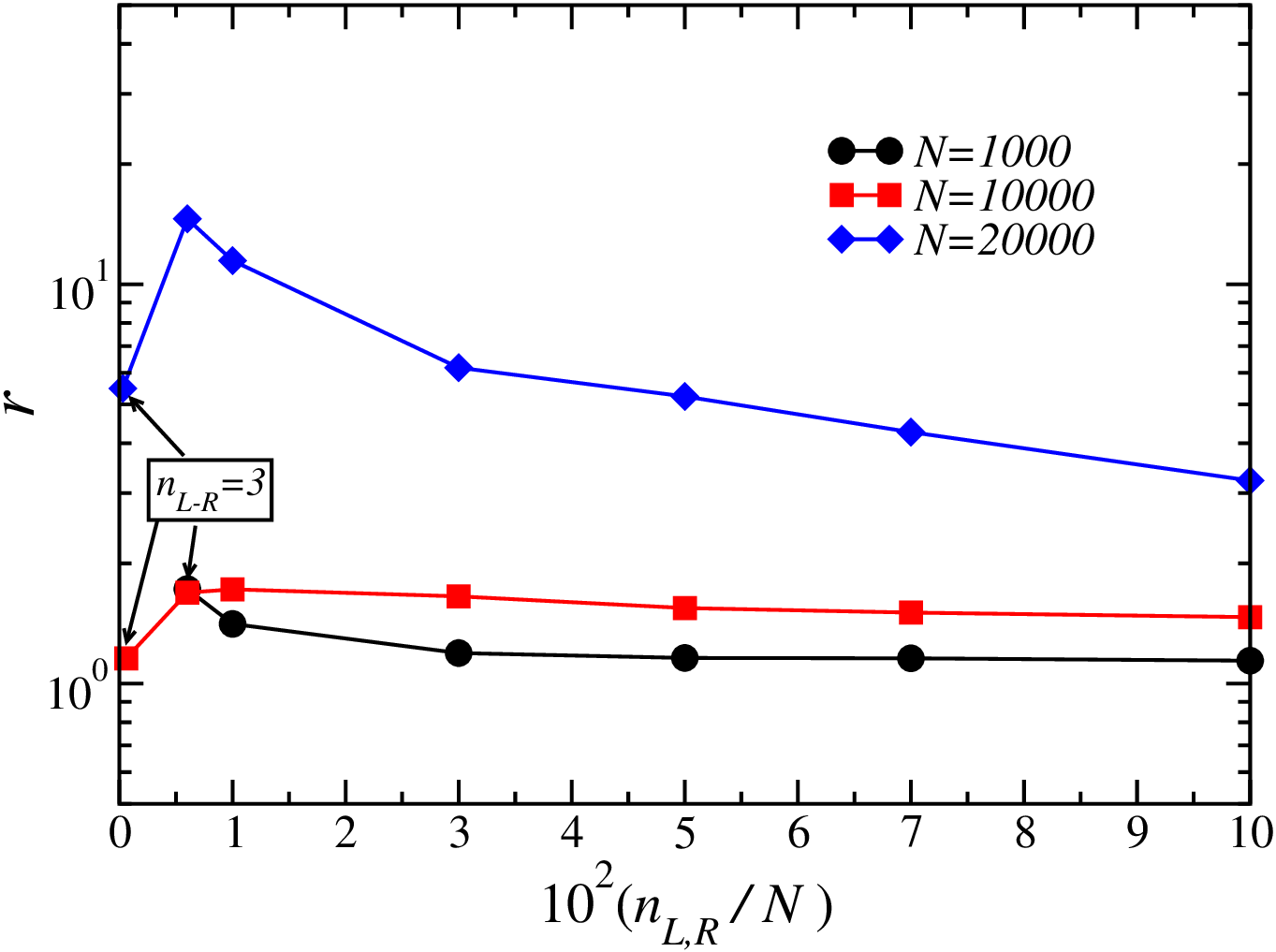}
\caption{(Color online) Thermal rectification for three system sizes ($N=1000$, $10\,000$, and $20\,000$), all with $\rho=0.1$, vs percentage of thermostated oscillators (view text for more details). Arrows indicate previously obtained values for the case $n_{_{L,R}}=3$.}
\label{fig:cinco}
\end{figure}

\begin{figure}
\includegraphics[width=0.99\linewidth,angle=0.0]{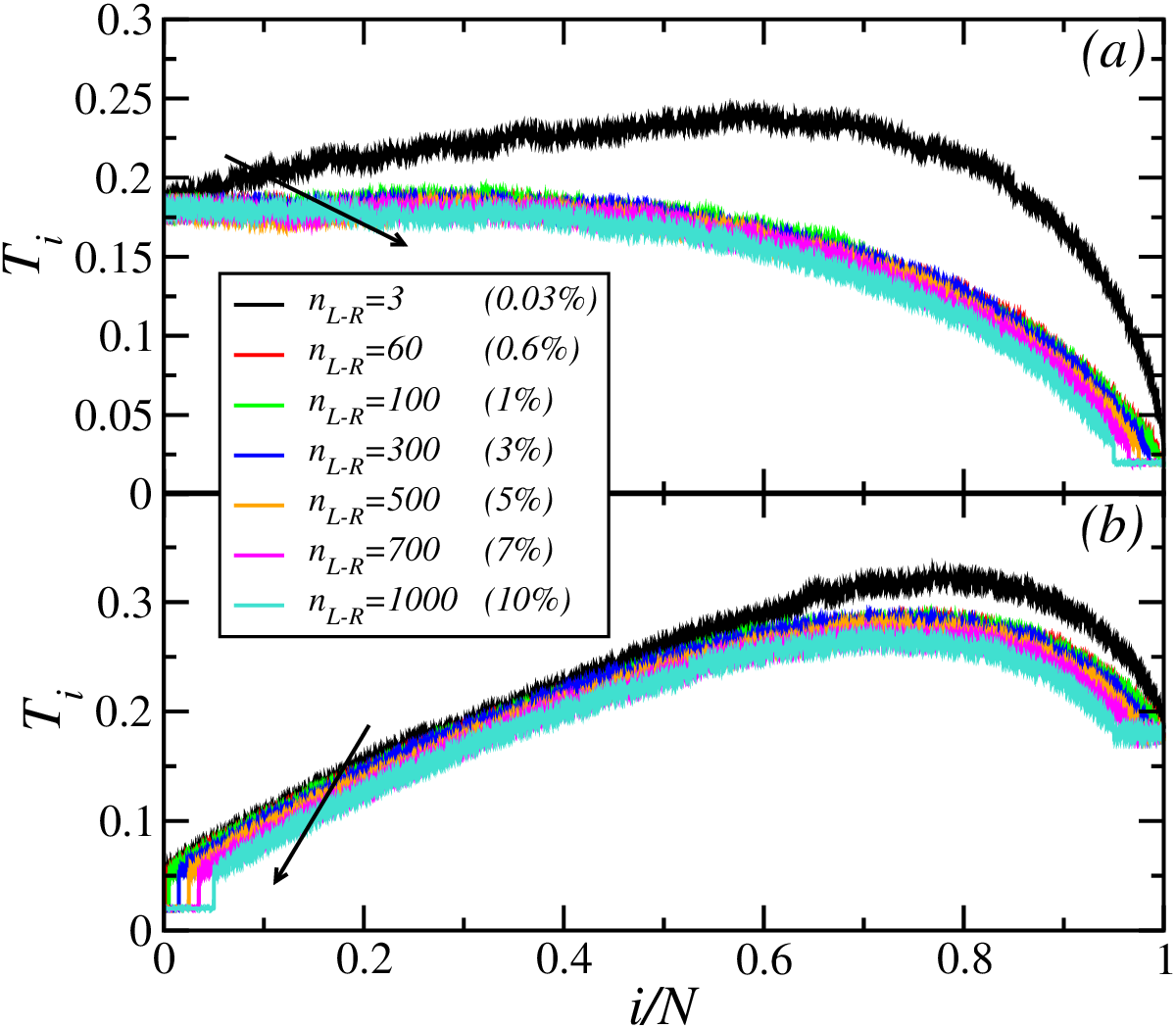}
\caption{(Color online) Temperature profiles for cases (a) $J_{+}$ and (b) $J_{-}$, both with $\rho=0.1$ and $N=20\,000$ of previous figure. Arrows indicate the direction in which the percentage of thermostated oscillators increases; the same $T_{0}$ and $\Delta T$ values are used as in Fig.~\ref{fig:uno}. }
\label{fig:seis}
\end{figure}

Now, just as determining the mass loading of the lattice by $\rho$ leads to an increase in rectification, it would seem natural to explore if that same idea could be applied to other structural variables of the system with that same aim. Therefore we performed some further simulations, but now taking $n_{_{L,R}}/N=$const., thus scaling the $n_{_{L,R}}$ value as $N$ increases. The results for three different system sizes are plotted in Fig.~\ref{fig:cinco}. It is clear that there is some variation of $r$ for a small fraction of thermostated oscillators, but as the latter increases, rectification approaches a constant value that increases as the lattice size does so, although at a much smaller rate compared to the increase obtained considering $\rho$ in Fig.~\ref{fig:tres}(b). Furthermore, for $N=20\,000$ there seems to be evidence that there is an asymptotic decay of $r$ for an even larger fraction of thermostated oscillators, although further simulations would be necessary to corroborate this last assertion. Nevertheless, we consider that the increase in rectification efficiency already obtained is not large enough to justify the effort. The TPs for the points of the $N=20\,000$ curve in Fig.~\ref{fig:cinco} are displayed in Fig.~\ref{fig:seis}(a) and ~\ref{fig:seis}(b) for $J_+$ and $J_-$, respectively. It is clear that the local temperature at each lattice site becomes increasingly independent of the $n_{_{L,R}}/N$ ratio as this latter increases. We readily observe, for the $J_+$ case in Fig.~\ref{fig:seis}(a), that the local temperature fluctuates around a constant value in most of the first half of the lattice, a feature which certainly accounts for the rectification value $<10$ reported in Fig.~\ref{fig:cinco} for this system size.

\section{concluding remarks\label{sec:Disc}}

One of the most severe problems encountered in systems that display thermal rectification is that this property disappears as the system size increases. Important advances have been achieved, although with a substantially different type of asymmetry than the one herein considered and based on macroscopic mechanisms~\cite{Hu06b}. The low rectification efficiency of the mass-graded FPU lattice is certainly one of the reasons this system has not recently received much attention, although it was inspired in the first experimental implementation of a rectifier, as already mentioned. Naively it would be reasonable to expect that the rectifying effects of the mass gradient would disappear as the system size increases with $\Delta M$ held constant. Surprisingly, the rectification has a modest increase at large $N$ values which has its origin in the peculiar shape of the TP not encountered in previous studies of this system~\cite{Yang07,Romero13}. It is important to mention that the uncommon behavior of the TP, i.e., the central region of the lattice having a higher temperature than either of the heat reservoirs, was first observed when the so-called FPU-$\alpha\beta$ lattice was subjected to the influence of time-varying thermal reservoirs with the same average temperature but an instant temperature oscillating at different frequencies; in that case a clear-cut connection between the symmetrical local temperature profile and the obtained ratcheting heat flux could not be detected~\cite{Nianbei09}. A non-monotonic TP was also obtained in a rotator lattice subjected to a constant mechanical forcing at one of its ends~\cite{Iacobucci11}. In this last system local equilibrium holds for sufficiently long lattices and, when the mechanical forcing is strong enough, the heat flux can be increased by an inverse temperature gradient. Furthermore, it has been known for some time that inside a harmonic lattice with certain type of interactions the direction of the heat flux cannot (in general) be supposed from the temperature gradient~\cite{Eckmann04}. The aforementioned rotator system affords an example of an anharmonic lattice where a seemingly counterintuitive direction of the heat flux can be observed. Furthermore, when an on-site FK potential is included, together with the original nearest-neighbour FPU one, in the potential energy, it has been observed that the heat flux is not constant and equal for each site, but increases linearly, as already mentioned~\cite{Zhao09}. Therefore, with the mass being a local, on-site variable, it can be considered that the mass density gradient plays a similar role to that of the aforementioned FK potential in creating the aforementioned characteristic feature of the heat flux profile. Nevertheless, its effect is more complex, since it induces two opposing heat fluxes on the lattice by means of the asymmetry of the TP. Our work goes beyond previously mentioned ones since an apparently counterintuitive heat flux not only is observed, but has actually been employed to increase the rectification of the mass-graded FPU lattice.

The introduced mass gradient density $\rho$ presents a number of advantages for controlling rectification in this system. From the experimental point of view one cannot guarantee that the considered system (nanotube-nanowire) of different system sizes fabricated separately are identical. However, it would be feasible to control the mass loading of a fixed-sized system by the herein presented method. Furthermore, since the scaling exponent $\alpha$ in the relation $r\sim\exp(\alpha N)$  presents two regions in its $\rho$ dependence, we afford a method to control the rectification that becomes largely independent of the system size $N$.

Since our proposed mass-loading scheme depends on a mass distribution along the length of the system and not on a fixed amount of it, the proposed parameter could be more easily controlled in an experimental situation. Since in the experimental study with nanotubes only half of the system was mass-loaded~\cite{Chang06a}, it remains to determine in which way our results are modified if such mass-loading scheme is employed, among many other possibilities. Furthermore, based on the reported evidence we have argued that the observed rectification increase can be easily explained from the asymmetric structure of the TP that results from the mass-loading. We have to emphasize that this connection is far from obvious since, as mentioned above, even for the problem of heat conduction no simple relation could be obtained between a non-monotonic TP and energy transport properties of the system~\cite{Zhao09,Iacobucci11,Nianbei09}. Finally, we remark that our study could be relevant to the problem of thermal rectification in higher dimensions, wherein much larger $r$ values can be obtained, but that rapidly diminishes as the system size increases, both in two-\cite{Lan06} and three-dimensional systems~\cite{Lan07}.

\smallskip
\begin{acknowledgments}
M.~R.~B. thanks CONACyT, M\'exico, and A.~G.~A. thanks ``Programa Institucional de Formaci\'on de Investigadores" I.P.N., M\'exico, for financial support. The authors also thank Federico Vazquez-Hurtado and Fernando Salazar-Posadas for useful comments and discussions. AMDG.
\end{acknowledgments}


\bibliographystyle{prsty}

\end{document}